\documentclass[]{spie}  

\usepackage{amsmath,amsfonts,amssymb}
\usepackage{graphicx}
\usepackage{caption}
\usepackage{subcaption}
\usepackage[colorlinks=true, allcolors=blue]{hyperref}

\title{A 3D-Printed Table for Hybrid X-ray CT and Optical Imaging of a Live Mouse}

\author[a]{Wenxuan Xue}
\author[a]{Yuxuan Liang}
\author[a]{Mengzhou Li}
\author[a]{Shan Gao}
\author[a]{Xavier R. Intes}
\author[a]{Ge Wang}
\affil[a]{Department of Biomedical Engineering, Rensselaer Polytechnic Institute, USA 12180}

\authorinfo{For further information, please send correspondence to G. Wang.\\ G. Wang: ge-wang@ieee.org}

\begin{document} 
\maketitle

\begin{abstract}
Multimodal imaging has shown great potential in cancer research by concurrently providing anatomical, functional, and molecular information in live, intact animals. During preclinical imaging of small animals like mice, anesthesia is required to prevent movement and improve image quality. However, their high surface area-to-body weight ratio predisposes mice, particularly nude mice, to hypothermia under anesthesia. To address this, we developed a detachable mouse scanning table with heating function for hybrid x-ray and optical imaging modalities, without introducing metal artifacts. Specifically, we employed Polylactic Acid (PLA) 3D printing technology to fabricate a customized scanning table, compatible with both CT and optical imaging systems. This innovation enables seamless transportation of the table between different imaging setups, while its detachable design facilitates maintaining a clutter-free operational environment within the imaging systems. This is crucial for accommodating various projects within the same scanner. The table features positioned fixation points to secure mice, ensuring positional consistency across imaging modalities. Additionally, we integrated a carbon nanotube-based heating pad into the table to regulate the body temperature of mice during examinations, providing an ethical and effective temperature maintenance solution. Our evaluations confirmed the table’s ability to maintain a 30g water bag at approximately 40℃, effectively regulating mouse body temperature to an optimal 36℃ during preclinical imaging sessions. This scanning table serves as a useful tool in preclinical cancer research, offering a versatile tool that upholds animal welfare standards.  
\end{abstract}


\section{INTRODUCTION}
\label{sec:intro}  

Multimodal imaging has great potential in preclinical research because it allows the concurrent acquisition of anatomical, functional, and molecular information in live and intact animals \cite{pian2014multimodal}. One powerful combination is in vivo imaging using X-ray computed tomography (CT) alongside optical imaging, which together can provide enhanced and complementary insights into the progression and treatment of complex diseases, such as cancer \cite{herranz2012optical} that often exhibit highly intricate spatiotemporal characteristics, making it vital to leverage the strengths of both imaging modalities. While CT is highly effective for visualizing the geometry and anatomy of animal tissues, as well as for detecting contrast materials, optical imaging excels in delivering functional information, such as tissue oxygenation and metabolic activity.

A critical consideration during such imaging procedures is ensuring that the animal's position remains stable and consistent across different imaging modalities. To achieve this, it is advantageous to use a scanning table that can be easily transported between systems, allowing the animal to remain in situ throughout the imaging process and facilitating straightforward image registration across different imaging platforms.

Regardless of the system-specific table designs, one essential aspect is the need to maintain the animal's body temperature during the imaging process. Anesthesia, commonly used in preclinical imaging, often leads to a significant drop in body temperature, especially in small rodents. They are prone to rapid heat loss due to their high surface area-to-volume ratio \cite{Gargiulo_2012}. In addition, animals can experience further heat loss due to the typically cool environments of imaging rooms, the scanners themselves, and the delivery of cool anesthetic carrier gases. Without active heating, the combination of these factors can quickly lead to hypothermia, which may impact the physiology and thus the outcomes of the experiment.

Several methods for maintaining body temperature during imaging have been proposed, including circulating warm liquids or air \cite{suckow2009multimodality}\cite{navarro2021}, using pre-warmed heat pads, and employing metallic heating resistors \cite{caro2013}. Although each of these methods can offer thermal support, many of them are unsuitable for use with micro-CT or optical scanners. Large heating devices can be impractical due to space constraints in the imaging apparatus, while others can cause image distortion, particularly in CT scanners where metallic components can introduce serious artifacts. For example, circulating warm liquids or air requires additional space within the scanner and poses challenges due to continuous heat loss during delivery, necessitating a high power demand to maintain adequate temperature. Pre-warmed heat pads, while more compact, cannot provide consistent heat over the duration of the imaging process, as they gradually return to room temperature. Metallic heating pads, although capable of maintaining constant heat, pose a significant problem for CT imaging due to the generation of metal artifacts, which compromise image quality \cite{suetens}.

Here, we present a 3D-printed scanning table that can be transported between CT and optic imaging systems keeping position of mouse constant with heating system using carbon nanotube heater which presents no detectable image artifacts.
\begin{figure}[htp]
    \centering
    \begin{subfigure}[b]{0.45\textwidth}
        \centering
        \includegraphics[width=\textwidth]{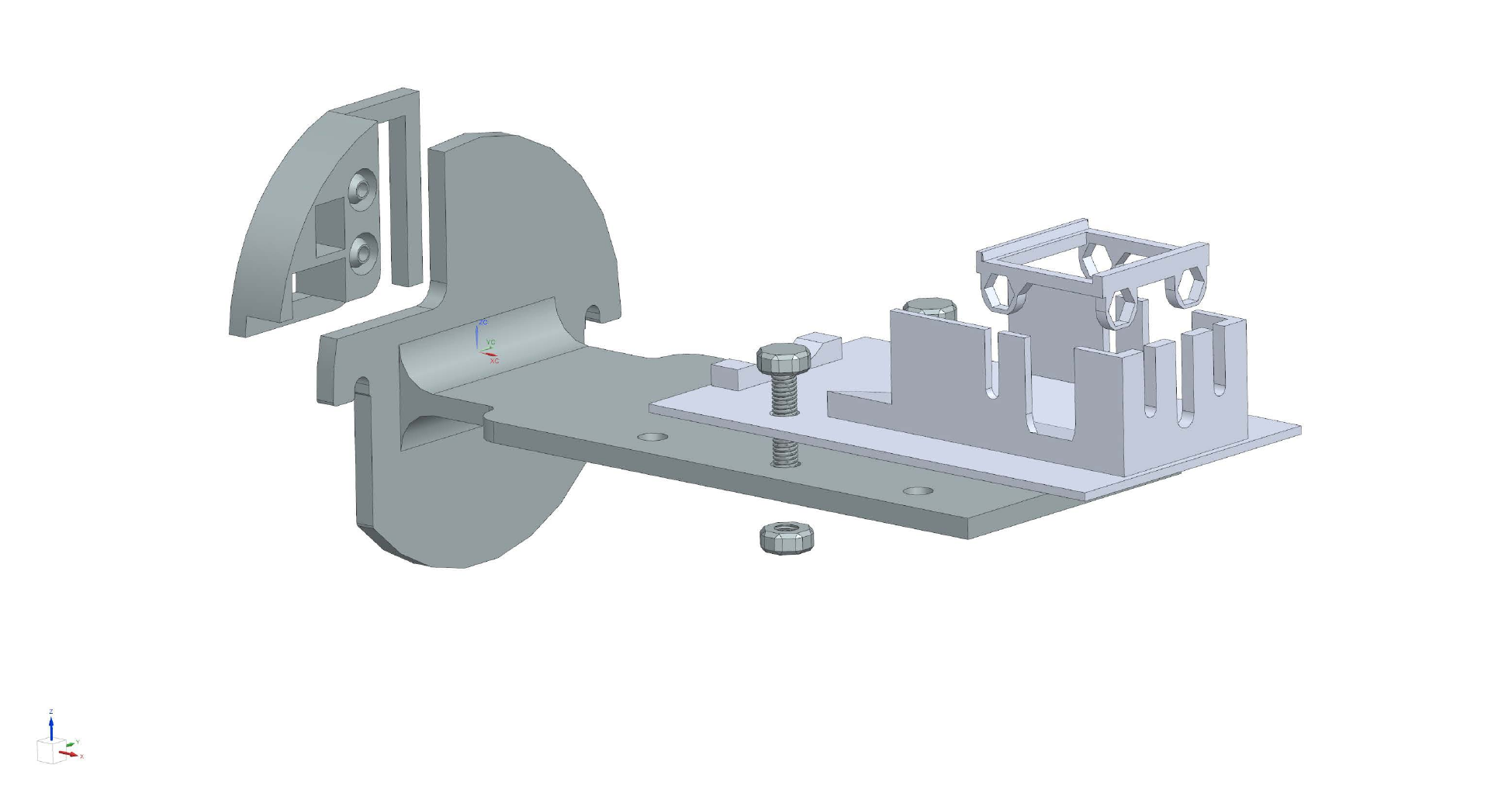}
        \caption{}
        \label{scatter}
    \end{subfigure}
    \hspace{5mm}
    \begin{subfigure}[b]{0.45\textwidth}
        \centering
        \includegraphics[width=\textwidth]{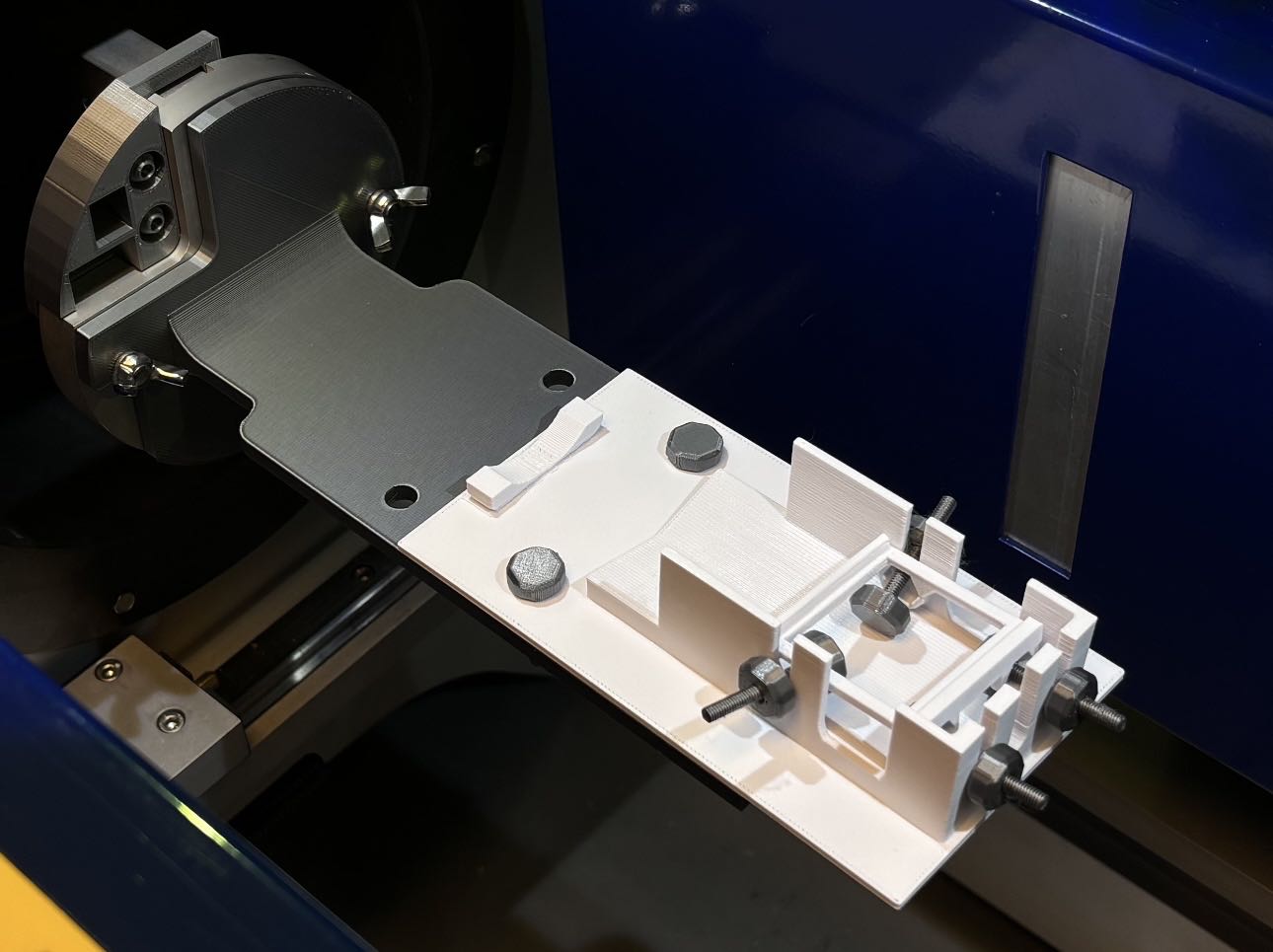}
        \caption{}
        \label{method}
    \end{subfigure}
\caption{Designed 3D-printed table. (a) Design sketch created using Siemens NX CAD software, and (b) photograph of the printed table set up in the MARS19 CT scanner. The model was sliced using PrusaSlicer software and 3D-printed using a Prusa Mk3S printer with PLA filament.}
    \label{fig1and2}
\end{figure}

\section{Materials and Methods}
\subsection{3D-printed Table}
To address the need for a compact table suitable for both micro-CT and optical imaging, we designed a mouse table in CAD and constructed it using Polylactic Acid (PLA) 3D printing technology. The table’s design includes a base that connects to the gantry of the scanner, a detachable bed, and a transparent cover. The base can be securely attached to the CT scanner gantry via an interface connector, which also accommodates wiring and tubing within the scanner.

The supporting bar holds the detachable bed, where the mouse is positioned during scanning. When switching from CT to optical imaging, the bed can be easily removed from the CT scanner and placed on a microscope. The transparent cover improves observation of the mouse's position and helps maintain its alignment during scanning. The dimensions of the table are compact—75mm in width, 30mm in height—with a cross-sectional area slightly larger than the average mouse size, ensuring short imaging durations. The length of the table is adjustable from 180mm to 260mm, allowing flexible positioning for various imaging needs. Furthermore, the base is designed to be easily detachable from the scanner gantry, enabling versatility across different imaging projects. Figure \ref{scatter} and \ref{method} show the conceptual and real-life images of the designed system.

\subsection{Carbon Nanotube Heater}

The heating system employs a flexible carbon nanotube-based electrothermal film heater, known for its high efficiency, minimal operational noise, and excellent safety profile. The system comprises three layers, with the central layer acting as the heating element, providing uniform heating across its surface. The outer layers serve as protective barriers, ensuring safe and effective operation. The heater operates at 5V and 10W, delivering safe, low-voltage performance, with no associated safety risks. The heating pad features a 99\% electrothermal conversion rate and emits heat in the far-infrared spectrum, which enhances its efficiency while ensuring safe and effective warming. With a lifespan of up to 50,000 hours, the system is designed for long-term, reliable use.

The heating pad's flexible nature allows it to conform to the shape and dimensions of the scanning bed. It is placed on the bottom surface of the bed, topped with a sponge mat that contours to the mouse's body, ensuring consistent positioning. An additional layer between the heating pad and the mouse prevents potential burns from overheating. The heating pad is powered via an extension cord connected at the gantry.

\section{Experiments and Results}
\subsection{Homeothermic Maintenance in Simulation Testing}
\begin{figure}[htp]
    \centering 
        \includegraphics[width=0.6\textwidth]{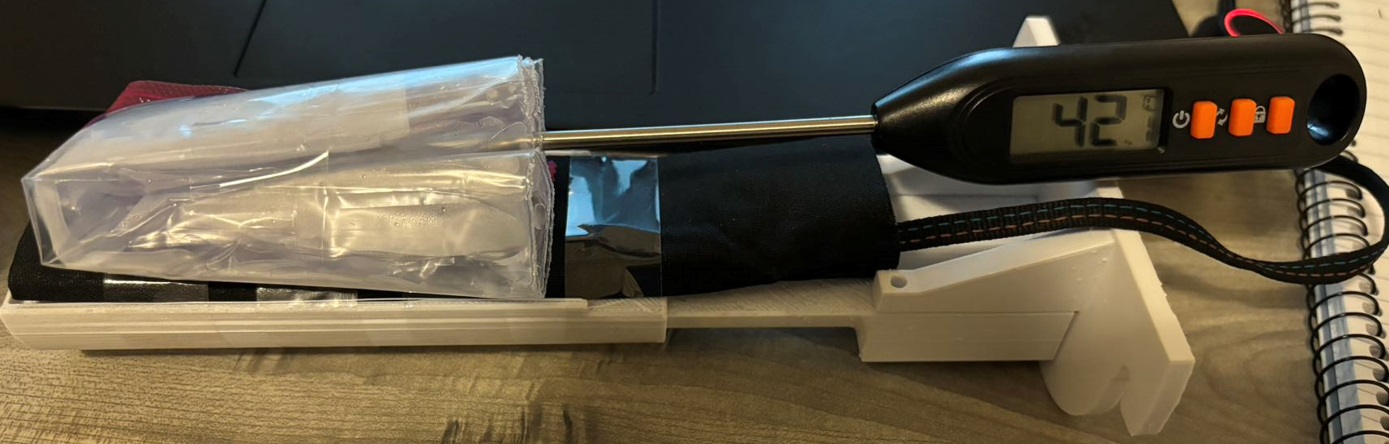}
      \vspace{5mm}
       \caption{Experimental setup for homeothermic maintenance testing.}
    \label{Expsetup_heat}
\end{figure}

To evaluate the heating pad's performance, surface temperature measurements were first taken using an infrared thermometer to assess the heating efficiency at steady state. Subsequently, a simulation test was conducted to assess its efficacy in maintaining homeothermic conditions. In this test, a 40g water bag, simulating the thermal mass of a mouse, was placed on the heating pad. Internal temperature measurements of the water bag were recorded at 5-minute intervals over a 180-minute period using a digital thermometer. The ambient room temperature was maintained at 24°C, with the initial temperature of the water bag also at approximately 24°C. Figure \ref{Expsetup_heat} shows the experimental setup for homeothermic maintenance testing.

\begin{figure}[htp]
    \centering 
        \includegraphics[width=0.7\textwidth]{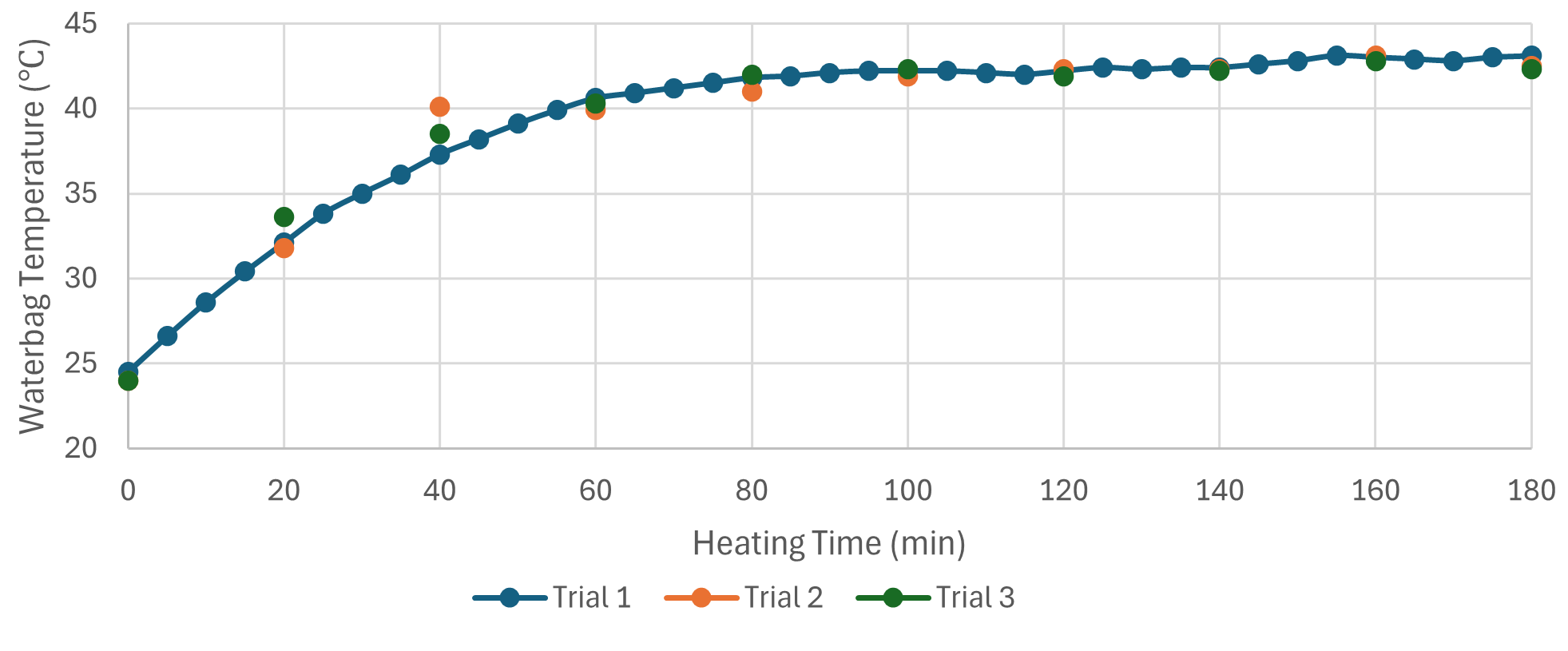}
      \vspace{5mm}
       \caption{Experimental setup for homeothermic maintenance testing.}
    \label{heat_results}
\end{figure}

Figure \ref{heat_results} illustrates the temperature profiles of the water bag during three independent trials. The core temperature of the water bag showed a steady increase over time. Within 20 minutes, the temperature rose from 24°C to over 30°C, demonstrating the system's efficiency in raising the temperature in a short period of time. The target temperature was set at 40°C, representing a typical homeothermic range \cite{hankenson2018effects}. This target was successfully reached within 45 minutes, indicating a consistent and controlled rate of temperature increase.

Once the target temperature of 40°C was achieved, the system maintained the water bag's temperature within a narrow range of 40°C ± 2°C for the next 135 minutes. This stability demonstrates the system's ability to maintain a homeothermic condition over an extended period, which is critical for ensuring consistent heating temperature during the imaging session. Most imaging period was intentionally kept under 3 hours, as the system was able to provide stable thermal maintenance throughout the entire duration.

\begin{figure}[htp]
    \centering 
        \includegraphics[width=0.5\textwidth]{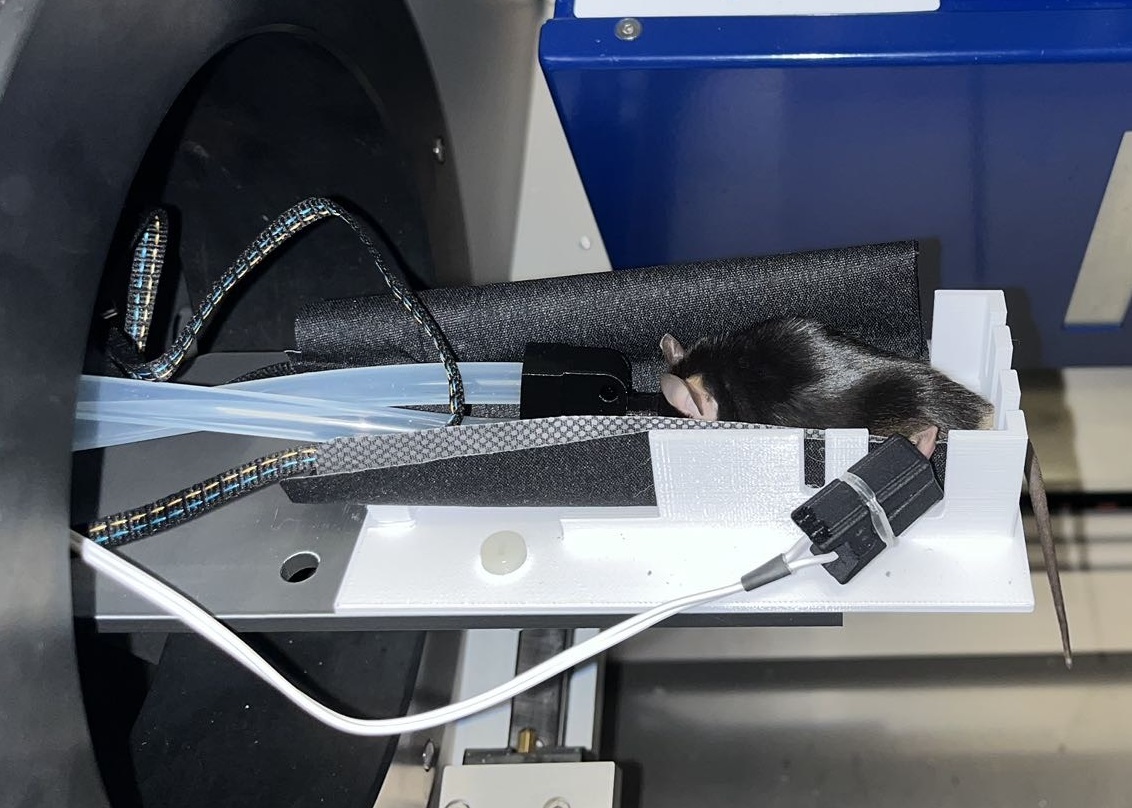}
      \hspace{5mm}
       \caption{Experimental setup for CT imaging and compatibility testing on a sacrificed mouse.}
    \label{Expsetup}
\end{figure}

\subsection{CT Imaging and Compatibility}
To validate the compatibility of the 3D-printed table with CT imaging, a sacrificed mouse was scanned using the MARS-19 photon-counting micro-CT scanner. Scans were performed at 118kVp and 12µA \cite{CT_para}, confirming that the table did not introduce any artifacts, and ensuring that it is suitable for hybrid imaging applications.

\begin{figure}[htp]
    \centering
    \begin{subfigure}[b]{0.3\textwidth}
        \centering
        \includegraphics[width=\textwidth]{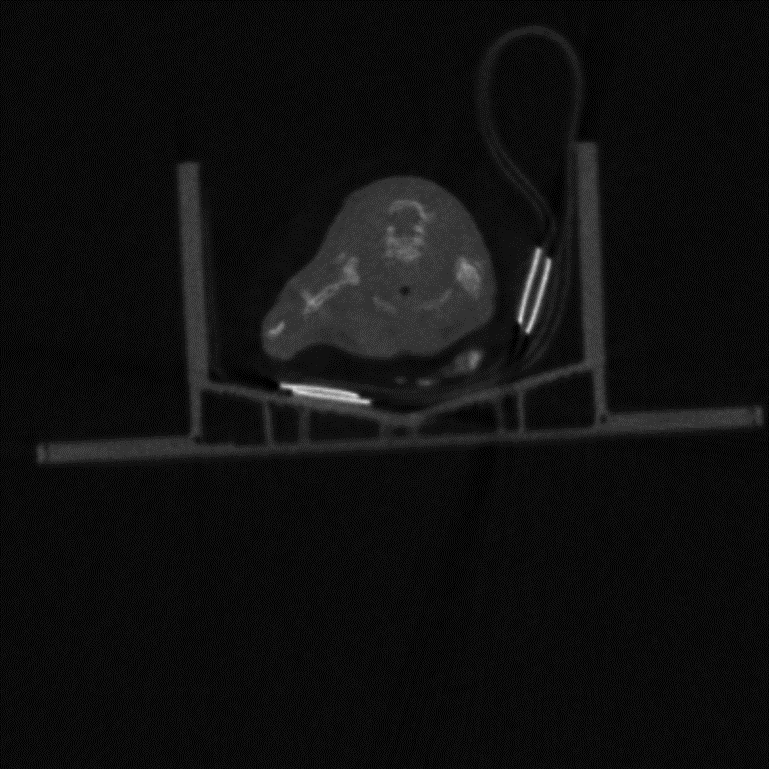}
        \caption{}
        \label{fig:sub1}
    \end{subfigure}
    \hspace{5mm}
    \begin{subfigure}[b]{0.3\textwidth}
        \centering
        \includegraphics[width=\textwidth]{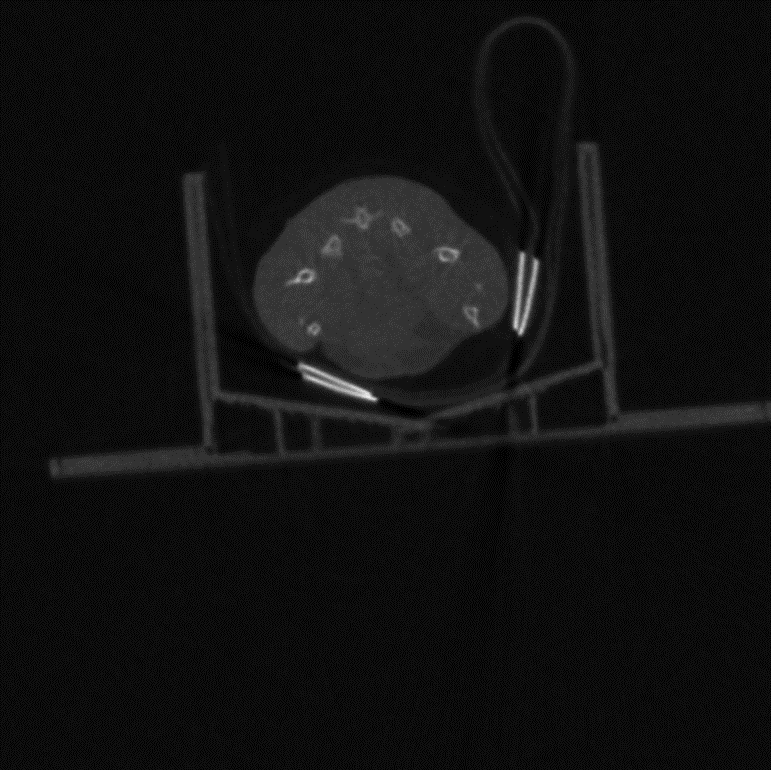}
        \caption{}
        \label{fig:sub2}
    \end{subfigure}
    
    \vspace{5mm}
    
    \begin{subfigure}[b]{0.3\textwidth}
        \centering
        \includegraphics[width=\textwidth]{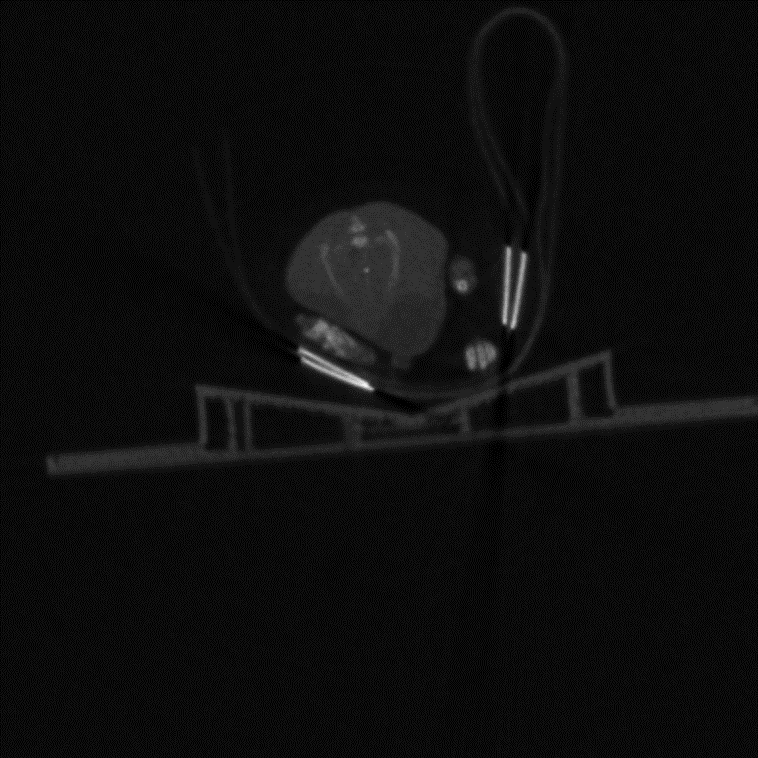}
        \caption{}
        \label{fig:sub3}
    \end{subfigure}
    \hspace{5mm}
    \begin{subfigure}[b]{0.3\textwidth}
        \centering
        \includegraphics[width=\textwidth]{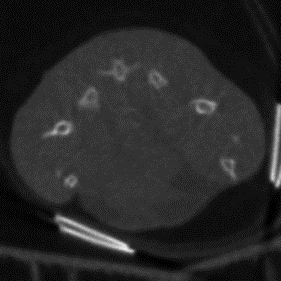}
        \caption{}
        \label{fig:sub4}
    \end{subfigure}
    
    \caption{Reconstructed CT images for compatibility testing. Selected cross-sectional images of the mouse: (a) neck, (b) chest, (c) leg, and (d) zoomed-in chest (image (a)). Window size: 121.60\,mm $\times$ 121.60\,mm.}
    \label{recon}
\end{figure}

In the reconstructed CT images (Figure \ref{recon}), some bright lines are observed, corresponding to the cross-sectional images of carbon nanotubes used in the heating pad beneath the mouse. These carbon nanotubes exhibit higher brightness than other components in the image, causing minor streak artefacts. However, these artefacts do not intersect with the mouse image and do not affect the visualization of the mouse’s anatomical structures. The high-quality imaging of the mouse allows for accurate assessment of internal anatomy without distortion, enabling precise measurements and observations throughout the experiment. This clarity in imaging further supports the reliability of the experimental setup for future applications requiring detailed anatomical assessments.

\section{Discussion}
We successfully developed a novel, customized, detachable scanning table that is compatible with both X-ray computed tomography (CT) and optical imaging systems. This design addresses several critical needs in preclinical imaging, particularly in the context of multimodal imaging where maintaining the animal’s position between systems and providing consistent thermal support are essential for reliable and repeatable results.

Our experiments confirmed that the integrated carbon nanotube-based heating system was able to effectively regulate temperature at 40°C for up to 3 hours, which exceeds the duration of typical imaging sessions. Maintaining consistent body temperature during preclinical imaging is crucial, as small animals are highly susceptible to hypothermia when anesthetized. The stable thermal environment provided by the heating pad ensures that the physiological conditions of the mouse remain as close to normal as possible during imaging, thereby minimizing any potential temperature-related physiological deviations that could affect experimental outcomes. This level of temperature regulation also highlights the system's potential for use in a range of preclinical studies where prolonged imaging sessions are required.

Another significance of our scanning table is its ability to function without introducing image distortions, particularly metal artifacts during CT imaging, which are commonly associated with traditional heating pads that utilize metallic heating resistors. The carbon nanotube heater in our system did not interfere with CT image quality, as confirmed by our experiments using the MARS-19 photon-counting micro-CT scanner. The absence of metal artifacts ensures high-quality imaging, preserving the anatomical details necessary for accurate assessments. The compatibility of the scanning table with both imaging modalities allows for seamless integration into hybrid imaging setups, ensuring that data from different modalities can be accurately registered and compared.

In addition to its functional advantages, the 3D-printed scanning table offers integration with existing preclinical CT systems. Designed with adjustable dimensions and mounting points, the CAD model can be customized to fit a variety of scanners beyond the MARS19 model used in our study. Installation is straightforward and requires no additional hardware, thus minimizing setup time and technical complications. Constructed from lightweight and radiolucent PLA filament, the table does not interfere with imaging quality and facilitates easy reproduction and customization using standard 3D printers. Importantly, the table does not necessitate any changes to scanner calibration or imaging software, allowing existing imaging protocols to remain unaltered and preserving consistency in experimental results. An consideration is the potential need for extension cords or power adapters to connect auxiliary equipment like the heating pad and physiological monitoring devices within the scanner environment.

Future work involves applying this system in live in-vivo preclinical hybrid X-ray and optical imaging of mice, where this detachable scanning table will be employed to monitor complex diseases such as cancer. These in-vivo studies will be instrumental in demonstrating the full utility of the scanning table and its heating system. Additionally, further validation of the system’s performance across different imaging modalities and experimental conditions will help solidify its role in multimodal imaging setups. Future developments may also explore enhancements to the table’s design, such as optimizing its dimensions for different animal models or integrating advanced materials for improved thermal insulation. One potential avenue for improvement involves making the heating pad system more intelligent, incorporating real-time temperature monitoring from sensors attached to the mouse. Based on the continuous data gathered from these sensors, the heating pad could automatically adjust its heat output to maintain the mouse’s body temperature within an optimal range. This automated control system would help ensure consistent thermal regulation throughout the imaging session, preventing fluctuations that could impact physiological stability. Furthermore, expanding the table’s functionality to support more advanced vital sign monitoring or to integrate automated control systems for adjusting both the heating and anesthesia systems could further enhance its utility in preclinical research. These developments would not only improve the overall efficiency of the experimental process but also contribute to better animal welfare by ensuring precise and individualized temperature control during imaging.

\acknowledgments 
 
We would like to express our sincere gratitude to Antigone McKenna for providing the mice used in this study. This research would not have been possible without her valuable contributions. Additionally, we are deeply thankful for the funding support that made this work achievable. The financial assistance played a critical role in facilitating the research.

\bibliography{report} 
\bibliographystyle{spiebib} 

\end{document}